\documentclass[aps,prb,reprint,superscriptaddress,showpacs]{revtex4-1}

\usepackage{amsmath,amssymb,bm}
\usepackage{graphicx}

\begin{document}

\title{Geometrical Pumping in Quantum Transport: Quantum Master Equation Approach}

\author{Tatsuro Yuge}
\email[]{yuge@acty.phys.sci.osaka-u.ac.jp}
\affiliation{Department of Physics, Osaka University, 
Machikaneyama-Cho, Toyonaka, 560-0043, Japan}
\affiliation{Yukawa Institute for Theoretical Physics, Kyoto University, 
Kitashirakawa Oiwake-Cho, 606-8502, Japan}

\author{Takahiro Sagawa}
\affiliation{Yukawa Institute for Theoretical Physics, Kyoto University, 
Kitashirakawa Oiwake-Cho, 606-8502, Japan}
\affiliation{The Hakubi Center, Kyoto University, Yoshida Ushinomiya-Cho, 606-8302, Japan}

\author{Ayumu Sugita}
\affiliation{Department of Applied Physics, Osaka City University, 
Sugimoto, 558-8585, Japan}

\author{Hisao Hayakawa}
\affiliation{Yukawa Institute for Theoretical Physics, Kyoto University, 
Kitashirakawa Oiwake-Cho, 606-8502, Japan}

\date{\today}

\begin{abstract}
For an open quantum system, we investigate the pumped current 
induced by a slow modulation of control parameters 
on the basis of the quantum master equation and full counting statistics.
We find that the average and the cumulant generating function of the pumped quantity 
are characterized by the geometrical Berry-phase-like quantities in the parameter space, 
which is associated with the generator of the master equation.
From our formulation, we can discuss the geometrical pumping under the control of 
the chemical potentials and temperatures of reservoirs. 
We demonstrate the formulation by spinless electrons in coupled quantum dots. 
We show that the geometrical pumping is prohibited for the case of non-interacting electrons 
if we modulate only temperatures and chemical potentials of reservoirs, 
while the geometrical pumping occurs in the presence of an interaction between electrons.
\end{abstract}


\maketitle

\section{Introduction}

When a quantum system is slowly and periodically modulated by two or more control parameters 
such as gate voltages, a net number of particles can be transported per period of the modulation 
even in the absence of dc driving force (e.g., bias voltage).
This phenomenon is known as an adiabatic pumping. 
The adiabatic pumping has received much attention 
because of its possibilities for quantized charge transport \cite{Thouless,NiuThouless,AvronSeiler,Kouwenhoven_etal,Pothier_etal,Fuhrer_etal,Kaestner_etal,Chorley_etal, 
AleinerAndreev,AndreevKamenev,MakhlinMirlin}, 
spin pumping \cite{Watson_etal,Mucciolo_etal,Governale_etal,Cota_etal,RiwarSplettstoesser,Splettstoesser2008,Deus_etal}, 
and qubit manipulation \cite{BrandesVorrath}, 
which are difficult to achieve in conventional stationary transport. 
The original idea of the adiabatic pumping was proposed by Thouless \cite{Thouless}, 
where a pumped current of a closed quantum system is related to 
the Berry phase \cite{Berry} of the ground state of the Hamiltonian 
\cite{Thouless,NiuThouless,AvronSeiler}. 

Since then, the idea of the adiabatic pumping has been applied to mesoscopic quantum systems. 
Experimentally, the pumping in Coulomb blockade regime \cite{Kouwenhoven_etal,Pothier_etal,
TsukagoshiNakazato,Watson_etal,Fuhrer_etal,Buitelaar_etal,Kaestner_etal,Chorley_etal} 
and in open quantum systems \cite{Switkes_etal,Giazotto_etal} has been developed. 
Theoretically, the formulation based on the time-dependent scattering theory \cite{BTP}
has been established. \cite{Brouwer,Avron_etal,AndreevKamenev,MakhlinMirlin,MoskaletsButtiker2001,CremersBrouwer,
Entin-Wohlman_etal,MoskaletsButtiker2002,Mucciolo_etal,Governale_etal,Kashcheyevs_etal,MoskaletsButtiker2004}. 
According to this formulation, the average pumped current can be expressed 
by the Berry phase associated with the scattering matrix \cite{Avron_etal}.
The cumulant generating function of pumped current can also be described 
by geometrical quantities \cite{MakhlinMirlin}.
For this reason, the adiabatic pumping is referred to as the geometrical pumping.
This scattering matrix approach is applicable to systems where the interaction can be neglected 
or treated in the mean field level.
A recent theoretical interest in this field is to understand the effects of interaction in the system 
on the adiabatic pumping \cite{AleinerAndreev,Citro_etal,Aono,Brouwer_etal,Splettstoesser2005,SelaOreg,
SchillerSilva,Devillard_etal,FiorettoSilva,Hernandez_etal,Kashuba,Deus_etal,
RenzoniBrandes,BrandesVorrath,Cota_etal,Splettstoesser2006,Splettstoesser2008,Reckermann_etal,
Hiltscher_etal,RiwarSplettstoesser}.

Similar phenomena have been studied in stochastic systems described by classical master equation 
\cite{Parrondo,Usmani_etal,Astumian1,SinitsynNemenman2007a,SinitsynNemenman2007b,Astumian2,Jarzynski,
Ohkubo,RenHanggiLi,Chernyak_etal1,Chernyak_etal2}, 
which is referred to as the adiabatic stochastic pumping.
The pumped current in the classical stochastic pumping has also geometrical properties; 
the cumulant generating function of the pumped current is expressed by a Berry-phase-like quantity 
that is associated with the generator of the classical master equation.
We shall refer to this quantity as the Berry-Sinitsyn-Nemenman (BSN) phase 
\cite{SinitsynNemenman2007a,SinitsynNemenman2007b}. 

There have also been several works on the adiabatic pumping for quantum open systems 
described by quantum master equation (QME) 
\cite{RenzoniBrandes,BrandesVorrath,Cota_etal,Splettstoesser2006,Splettstoesser2008,
Reckermann_etal,Hiltscher_etal,RiwarSplettstoesser}. 
Although the results for pumped charge or spin in specific models were provided in those works, 
the geometrical formulae for the adiabatic pumping described by QME have not been discussed so far.

In this paper, we investigate the quantum adiabatic pumping on the basis of QME, 
and derive general formulae of the cumulant generating function and average of the pumped quantity. 
These formulae are geometrical and expressed by a quantum analogous of the BSN phase 
which is associated with the generator of the QME. 
In QME approach, we can treat interaction between particles beyond the mean field level 
in any of perturbative, non-perturbative, or exact methods, 
depending on the model and analysis. 
In any methods, we can apply these formulae as long as we employ the QME, 
since our theoretical framework is independent of the details of the system. 
Therefore these are useful for analyzing a variety of applications of 
the adiabatic pumping such as a qubit rotation in quantum dots \cite{BrandesVorrath} 
and a spin pumping \cite{Cota_etal,Splettstoesser2008,RiwarSplettstoesser}. 
We note that the QME approach is also suitable to analyze systems 
that include dissipations and decoherences \cite{BrandesVorrath}.

This paper is organized as follows.
In Sec.~\ref{sec:GeneralResults}~A, the QME approach for an open system is described. 
The full counting statistics in the QME approach is discussed in Sec.~\ref{sec:GeneralResults}~B. 
For the adiabatic pumping, the geometrical formulae (BSN phase expressions) 
for the cumulant generating function and average of the pumped quantity 
are derived in Sec.~\ref{sec:GeneralResults}~C.
In this formulation, the temperatures and chemical potentials of reservoirs are parts of the control parameters. 
This is in contrast to most of the conventional studies on the adiabatic pumping, 
where only the parameters in the system Hamiltonian or in the coupling with the reservoirs are considered 
(for scattering matrix approach in the presence of ac voltage, see Ref.~\onlinecite{MoskaletsButtiker2004}).
In Sec.~\ref{sec:Example}, we demonstrate our theory by the spinless electron transport in quantum dot systems.
For non-interacting cases with (Sec.~\ref{sec:Example}~A) and 
without (Sec.~\ref{sec:Example}~B) the rotating wave approximation (RWA), 
we find that no geometrical pumping occurs if we control 
only the temperatures and chemical potentials of reservoirs.
In contrast, the geometrical pumping occurs for an interacting electron system 
even when we control only the reservoir parameters (Sec.~\ref{sec:Example}~C).
Section \ref{sec:Conclusion} is devoted to the discussion and conclusion.
In Appendix~\ref{derivaton_GQME}, 
we describe the details of the QME in the framework of the full counting statistics.
In Appendix~\ref{derivation_BSN_RWA}, we show the detailed derivation of the result 
for non-interacting case with the RWA.
In Appendix~\ref{equivalence_RWA_NonRWA}, we verify the equivalence between without and within the RWA 
concerning the unit-time generating function of the current between the system and reservoirs in a steady state. 
In Appendix~\ref{sec:Riwar}, we investigate the consistency of the results calculated in our scheme 
with the results in Ref.~\onlinecite{RiwarSplettstoesser}.

\section{General results}\label{sec:GeneralResults}

\subsection{Setup}

\begin{figure}[bt]
\begin{center}
\includegraphics[width=0.95\linewidth]{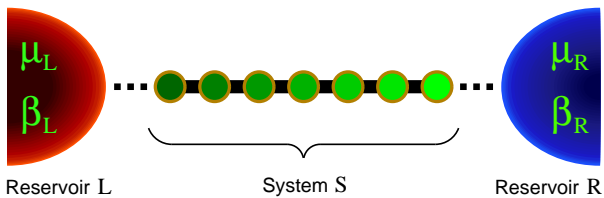}
\caption{(Color online) Illustration of our setup with two reservoirs (L and R).}
\label{illustration}
\end{center}
\end{figure}

We consider a quantum system S that is weakly coupled to reservoirs $\{{\rm R}_b\}$, 
where $b$ is an index of reservoirs (see Fig.~{\ref{illustration}} for a schematic).
The total Hamiltonian of the coupled system is 
$\hat{H}_{\rm tot} = \hat{H}_{\rm S} + \sum_b (\hat{H}_b + \hat{H}_{{\rm S}b})$, 
where $\hat{H}_{\rm S}$ is the system Hamiltonian, 
$\hat{H}_b$ is the Hamiltonian of the $b$th reservoir ${\rm R}_b$, 
and $\hat{H}_{{\rm S}b}$ is the interaction Hamiltonian between S and ${\rm R}_b$. 
If the interaction between the system S and the reservoirs is weak, 
the dynamics of S can be described by a QME 
for the reduced density matrix of S, which is denoted as $\hat{\rho}$.
Suppose that the initial state of the system S is decoupled with the reservoirs. 
Then, up to the second order in the system-reservoir coupling (Born approximation) 
with the Markov approximation \cite{BreuerPetruccione}, 
the QME for the system S reads 
\begin{align}
\frac{d\hat{\rho}(t)}{dt} = \mathcal{K} \hat{\rho}(t), 
\end{align}
where
\begin{align}
\mathcal{K} \hat{\rho} \equiv \frac{1}{i\hbar} & [\hat{H}_{\rm S} , \hat{\rho}] 
+ \sum_b \mathcal{D}_b \hat{\rho},
\\
\mathcal{D}_b \hat{\rho} 
\equiv - & \frac{1}{\hbar^2} \int_0^\infty dt' {\rm Tr}_b 
\bigl[ \hat{H}_{{\rm S}b} , [ \Check{H}_{{\rm S}b}(-t') , 
\hat{\rho} \otimes \hat{\rho}_b ] \bigr] .
\notag
\end{align}
Here the symbol ``~$\Check{~}$~'' stands for the interaction picture with respect to 
$\hat{H}_{\rm S} + \sum_b \hat{H}_b$, 
${\rm Tr}_b$ represents the trace over the $b$th reservoir, 
and $\hat{\rho}_b = e^{-\beta_b (\hat{H}_b -\mu_b \hat{N}_b)}/Z_b$ 
is the grandcanonical distribution 
with the inverse temperature $\beta_b$, chemical potential $\mu_b$, 
and the particle number operator $\hat{N}_b$ of the $b$th reservoir.
The time-evolution generator $\mathcal{K}$ of the QME depends on several parameters; 
the system parameters in $\hat{H}_{\rm S}$ and $\hat{H}_{{\rm S}b}$ 
such as the energy levels of quantum dots and the tunnel barriers between them, 
and the reservoir parameters, $\{\beta_b\}$ and $\{\mu_b\}$.
We write the set of these parameters as $\bm{\alpha}$.
The right eigenvalue equation for $\mathcal{K}$ is written as 
\begin{align}
\mathcal{K} \hat{\rho}_n(\bm{\alpha}) 
= \lambda_n(\bm{\alpha}) \hat{\rho}_n(\bm{\alpha}), 
\end{align}
where $\lambda_n(\bm{\alpha})$ is an eigenvalue of $\mathcal{K}$, 
$n$ is a label of the eigenvalues, 
and $\hat{\rho}_n(\bm{\alpha})$ is the corresponding right eigenvector.

By introducing the Hilbert-Schmidt inner product of linear operators $\hat{A}$ and $\hat{B}$ of the system S 
as ${\rm Tr}_{\rm S} (\hat{A}^\dag \hat{B})$, where ${\rm Tr}_{\rm S}$ is the trace over S, 
we define the adjoint $\mathcal{K}^\dag$ of the QME generator 
such that ${\rm Tr}_{\rm S} [(\mathcal{K}^\dag \hat{A})^\dag \hat{B}] 
= {\rm Tr}_{\rm S} (\hat{A}^\dag \mathcal{K} \hat{B})$ 
holds for any $\hat{A}$, $\hat{B}$.
We then have the left eigenvalue equation for $\mathcal{K}$: 
\begin{align}
\mathcal{K}^\dag \hat{\ell}_n(\bm{\alpha}) 
= \lambda_n^*(\bm{\alpha}) \hat{\ell}_n(\bm{\alpha}), 
\end{align}
where $\hat{\ell}_n(\bm{\alpha})$ is the left eigenvector corresponding to 
the eigenvalue $\lambda_n(\bm{\alpha})$.
In the following, we assume that $\mathcal{K}$ has the zero eigenvalue $\lambda_0=0$ 
without degeneracy, 
so that $\mathcal{K} \hat{\rho}_0=0$ and $\mathcal{K}^\dag \hat{\ell}_0=0$ hold.
This implies that the QME has a unique steady solution 
$\hat{\rho}_0(\bm{\alpha})$ 
for fixed $\bm{\alpha}$.
We note that $\hat{\ell}_0(\bm{\alpha})=\hat{1}$ (identity operator) 
holds for any $\bm{\alpha}$.

\subsection{Full counting statistics}

We consider the statistics of a quantity $\varDelta q$ transferred 
from the reservoirs to the system S during a time interval $\tau$.
The measurement scheme of $\varDelta q$ is as follows. 
First, at $t=0$ we perform a projection measurement of a reservoir variable $\hat{Q}$ 
to obtain a measurement outcome $q_0$. 
We assume $[\hat{Q},\hat{N}_b]=0$ and $[\hat{Q},\hat{H}_b]=0$ for any $b$.
For $t>0$, 
the system S undergoes the time evolution with interacting with the reservoirs. 
At $t=\tau$ we again perform a projection measurement of $\hat{Q}$ 
to obtain another measurement outcome $q_\tau$. 
Then $\varDelta q$ is defined as $\varDelta q = q_\tau - q_0$.

The cumulant generating function of the statistics is given by 
$S_\tau(\chi) = \ln \int P_\tau(\varDelta q) e^{i\chi \varDelta q} d\varDelta q$, 
where $P_\tau(\varDelta q)$ is the probability of $\varDelta q$ during $\tau$. 
$\chi$ is called the counting field, and the derivatives of $S_\tau(\chi)$ 
give the cumulants of $P_\tau(\varDelta q)$; e.g., 
$\langle \varDelta q \rangle_\tau = \partial S_\tau(\chi) / \partial (i\chi) |_{\chi=0}$.
Note that if $\hat{Q}$ is the $b$th reservoir's particle number $\hat{N}_b$ 
(Hamiltonian $\hat{H}_b$), 
then $\langle \varDelta q \rangle_\tau / \tau$ is the average 
of the particle (energy) current from the $b$th reservoir into the system S.

For calculating $S_\tau(\chi)$, we employ a method developed in the context of 
the full counting statistics \cite{EspositoHarbolaMukamel}.
In this method $S_\tau(\chi)$ is obtained from the solution of 
the modified equation of motion which is governed by the $\chi$-dependent Hamiltonian.
In the QME approach, this reads 
$S_\tau(\chi) = \ln {\rm Tr}_{\rm S} \hat{\rho}^\chi(\tau)$, 
where $\hat{\rho}^\chi(\tau)$ is a solution of the generalized QME (GQME): 
\begin{align}
\frac{d\hat{\rho}^\chi(t)}{dt} = \mathcal{K}_\chi \hat{\rho}^\chi(t).
\label{GQME}
\end{align}
Here the modified generator $\mathcal{K}_\chi$ is given by 
\begin{align}
\mathcal{K}_\chi \hat{\rho}^\chi 
&\equiv \frac{1}{i\hbar} [\hat{H}_{\rm S} , \hat{\rho}^\chi] 
+ \sum_b \mathcal{D}_b^\chi \hat{\rho}^\chi,
\label{GQMEgenerator}\\
\mathcal{D}_b^\chi \hat{\rho} &\equiv - \frac{1}{\hbar^2} \int_0^\infty dt' 
{\rm Tr}_b \bigl[ \hat{H}_{{\rm S}b} , [ \Check{H}_{{\rm S}b}(-t') , 
\hat{\rho} \otimes \hat{\rho}_b ]_\chi \bigr]_\chi ,
\notag
\end{align}
where $[\hat{O} , \hat{P}]_\chi \equiv \hat{O}^\chi \hat{P} - \hat{P} \hat{O}^{-\chi}$ 
and $\hat{O}^\chi \equiv e^{-i\chi\hat{Q}/2} \hat{O} e^{i\chi\hat{Q}/2}$.
See Appendix~\ref{derivaton_GQME} for details. 

We denote the left and right eigenvectors of $\mathcal{K}_\chi$ 
(for fixed $\bm{\alpha}$) corresponding to the eigenvalue $\lambda^\chi_n(\bm{\alpha})$ 
as $\hat{\ell}^\chi_n(\bm{\alpha})$ and $\hat{\rho}^\chi_n(\bm{\alpha})$, respectively.
They are normalized as 
${\rm Tr}_{\rm S} (\hat{\ell}^{\chi\dag}_m \hat{\rho}^\chi_n) = \delta_{mn}$.
We assign the label for the eigenvalue with maximum real part to $n=0$.
Then $\hat{\rho}^\chi(\tau) \sim e^{\lambda^\chi_0 \tau}$ for large $\tau$, 
which results in $\lim_{\tau\to\infty} S_\tau(\chi)/\tau = \lambda^\chi_0$.
Thus $\lambda^\chi_0(\bm{\alpha})$ 
is the unit-time cumulant generating function 
of the steady state for fixed $\bm{\alpha}$.\cite{EspositoHarbolaMukamel}
Note that if we set $\chi=0$, the GQME reduces to the original QME, and 
$\hat{\ell}^\chi_0$ and $\hat{\rho}^\chi_0$ to 
$\hat{\ell}_0=\hat{1}$ and the steady state $\hat{\rho}_0$, respectively.


\subsection{Geometrical pumping}

We slowly modulate the parameters $\bm{\alpha}$ along a curve $\mathcal{C}$ 
in the parameter space during a time interval $\tau$. 
If the system is in the instantaneous steady state 
for the value of $\bm{\alpha}_t$ at each time $t$ in the whole of the process, 
the cumulant generating function for $\varDelta q$ for this process 
is equal to the time integral of the unit-time cumulant generating function 
$\lambda^\chi_0(\bm{\alpha}_t)$ of the instantaneous steady state. 
In general, however, there exists additional (pumped) contribution; 
\begin{align}
S_\tau(\chi) = \int_0^\tau dt \lambda^\chi_0( \bm{\alpha}_t ) + S^{\rm ex}_\tau(\chi).
\end{align}
We call the latter contribution the excess part.
The excess part is intrinsic in the transitions between the steady states, 
and is of our interest.

We here derive the geometrical expression of the excess part of the generating function 
by using the method similar to those 
in Refs.~\onlinecite{SinitsynNemenman2007a,SagawaHayakawa}. 
First, to solve the GQME for a given curve $\mathcal{C}$ of $\bm{\alpha}$, we expand $\hat{\rho}^\chi(t)$ as 
\begin{align}
\hat{\rho}^\chi(t) 
= \sum_n c_n(t) e^{\Lambda^\chi_n(t)} \hat{\rho}^\chi_n (\bm{\alpha}_t),
\end{align}
where $\Lambda^\chi_n(t) \equiv \int_0^t dt' \lambda^\chi_n(\bm{\alpha}_{t'})$.
Inserting this equation into Eq.~(\ref{GQME}) 
and taking the Hilbert-Schmidt inner product 
with $\hat{\ell}^\chi_0(\bm{\alpha}_t)$, 
we obtain 
\begin{align}
\frac{dc_0(t)}{dt} = - \sum_n c_n(t) e^{\Lambda^\chi_n(t) - \Lambda^\chi_0(t)} 
{\rm Tr}_{\rm S}\left( \hat{\ell}^{\chi\dag}_0(\bm{\alpha}_t) 
\frac{d\hat{\rho}^\chi_n(\bm{\alpha}_t)}{dt} \right).
\label{eq_c0}
\end{align}
Now we assume the adiabatic condition, i.e., 
the modulation of the parameters $\bm{\alpha}$ is sufficiently slower 
than all the characteristic time scales of the system S. 
In many cases, the relaxation time $\tau_{\rm rlx}$, which is determined by the coupling with the reservoirs, 
is the longest time scale of S, so that the adiabatic condition should read 
(modulation time scale) $\gg \tau_{\rm rlx}$. 
[We will numerically confirm the necessity of this condition in Fig.~\ref{fig:DQD}(d).] 
Under this condition, we can approximate the sum on the right hand side of Eq.~(\ref{eq_c0}) by 
the contribution only from the term with $n=0$.
By solving this adiabatic approximation equation we obtain 
\begin{align}
c_0 (\tau) = c_0(0) 
\exp\left[ - \int_{\mathcal C} {\rm Tr}_{\rm S} 
\bigl( \hat{\ell}^{\chi\dag}_0 (\bm{\alpha}) 
d \hat{\rho}^\chi_0(\bm{\alpha}) \bigr) \right], 
\end{align}
where $d \hat{\rho}^\chi_0(\bm{\alpha}) \equiv d \bm{\alpha} \cdot 
\partial \hat{\rho}^\chi_0(\bm{\alpha}) / \partial \bm{\alpha}$.
If the initial state of the system S is the steady state with $\bm{\alpha}_0$, 
$\hat{\rho}^\chi(0) = \hat{\rho}_0(\bm{\alpha}_0)$, 
then $c_0(0) = {\rm Tr}_{\rm S}\Bigl[ \hat{\ell}^{\chi\dag}_0(\bm{\alpha}_0) 
\hat{\rho}_0(\bm{\alpha}_0) \Bigr]$.
We again use the adiabatic approximation to obtain 
\begin{align}
\hat{\rho}^\chi(\tau) &\simeq 
c_0 (\tau) e^{\Lambda^\chi_0(\tau)} \hat{\rho}^\chi_0 (\bm{\alpha}_\tau)
\notag\\
&= e^{\Lambda^\chi_0(\tau)} \hat{\rho}^\chi_0(\bm{\alpha}_\tau) 
{\rm Tr}_{\rm S} \Bigl( \hat{\ell}^{\chi\dag}_0(\bm{\alpha}_0) 
\hat{\rho}_0(\bm{\alpha}_0) \Bigr)
\notag\\
&\times \exp\left[ - \int_{\mathcal C} {\rm Tr}_{\rm S} \bigl( \hat{\ell}^{\chi\dag}_0(\bm{\alpha}) 
d \hat{\rho}^\chi_0(\bm{\alpha}) \bigr) \right] . 
\end{align}
Thus we obtain the excess cumulant generating function 
$S^{\rm ex}_\tau(\chi) = S_\tau(\chi) - \Lambda^\chi_0(\tau)$ for the slow modulation: 
\begin{align}
S^{\rm ex}_\tau(\chi) 
&= - \int_{\mathcal C} {\rm Tr}_{\rm S} \Bigl( \hat{\ell}_0^{\chi\dag}(\bm{\alpha}) 
d \hat{\rho}_0^\chi(\bm{\alpha}) \Bigr) 
\notag\\
&+ \ln {\rm Tr}_{\rm S} \Bigl( \hat{\ell}_0^{\chi\dag}(\bm{\alpha}_0) 
\hat{\rho}_0(\bm{\alpha}_0) \Bigr)
+ \ln {\rm Tr}_{\rm S} \hat{\rho}_0^{\chi}(\bm{\alpha}_\tau) . 
\label{CGF_ex}
\end{align}
This implies that $S^{\rm ex}_\tau(\chi)$ depends not on time $\tau$ 
but only on the curve $\mathcal{C}$ along which the parameters are varied.
The right-hand side of Eq.~(\ref{CGF_ex}) is analogous to 
the Berry phase in quantum mechanics, where $\hat{\ell}_0$ and $\hat{\rho}_0$ 
are both replaced by the eigen wave function of the Schr\"odinger equation.
We also note that $\Lambda^\chi_0(\tau)$ corresponds to the dynamical phase.

By differentiating Eq.~(\ref{CGF_ex}) with respect to $i\chi$, 
we obtain a geometrical expression of 
the average excess in the quantity $\varDelta q$:
\begin{align}
\langle \varDelta q \rangle^{\rm ex}_\tau 
= - \int_{\mathcal C} {\rm Tr}_{\rm S} \left( \hat{\ell}_0^{\prime\dag}(\bm{\alpha}) 
\frac{\partial \hat{\rho}_0}{\partial \bm{\alpha}}(\bm{\alpha}) \right) 
\cdot d\bm{\alpha}, 
\label{average_ex}
\end{align}
where $\hat{\ell}_0^\prime 
\equiv \partial \hat{\ell}_0^\chi / \partial(i\chi) |_{\chi=0}$. 
Equations (\ref{CGF_ex}) and (\ref{average_ex}) are regarded as quantum versions of 
the Berry-Sinitsyn-Nemenman (BSN) phases for the cumulant generating function and average, respectively, 
in slow parametric modulation \cite{SinitsynNemenman2007a,RenHanggiLi}.
We denote the integrand in Eq.~(\ref{average_ex}) as 
\begin{align}
\bm{A}(\bm{\alpha}) \equiv {\rm Tr}_{\rm S} \left( \hat{\ell}_0^{\prime\dag}(\bm{\alpha}) 
\frac{\partial \hat{\rho}_0}{\partial \bm{\alpha}}(\bm{\alpha}) \right), 
\end{align}
and refer to as the BSN vector potential for the average excess quantity.

Equality (\ref{average_ex}) implies that a finite net quantity of $\varDelta q$ 
can be transferred to the system S for a slow cyclic modulation of the parameters 
(i.e., for the case where the curve $\mathcal{C}$ is a closed loop)
even if there is no dc driving force such as temperature and chemical potential differences.
This is an adiabatic pumping.
For a cyclic process, by the Stokes theorem, Eq.~(\ref{average_ex}) is rewritten as 
\begin{align}
\langle \varDelta q \rangle^{\rm ex}_{\tau_c} 
= - \int_{\mathcal{S_C}} \sum_{m,n}
\frac{1}{2} F_{\alpha_m \alpha_n} d \alpha_m \wedge d \alpha_n , 
\label{average_ex_cycle}
\end{align}
where $\wedge$ is the wedge product, 
$\mathcal{S_C}$ is a surface enclosed by $\mathcal{C}$, and 
\begin{align}
F_{\alpha_m \alpha_n} \equiv 
{\rm Tr}_{\rm S} \left( \frac{\partial \hat{\ell}_0^{\prime \dag}}{\partial \alpha_m} 
\frac{\partial \hat{\rho}_0}{\partial \alpha_n}
- \frac{\partial \hat{\ell}_0^{\prime \dag}}{\partial \alpha_n} 
\frac{\partial \hat{\rho}_0}{\partial \alpha_m} \right). 
\label{BSN_curvature}
\end{align}
We refer to Eq. (\ref{BSN_curvature}) as the BSN curvature for average pumped quantity.

For some cyclic processes, 
the excess (pumped) quantity $\langle \varDelta q \rangle^{\rm ex}_{\tau_c}$ vanishes.
A sufficient condition for the ``no-pumping'' is that 
$F_{\alpha_m \alpha_n} = 0$ holds in $\mathcal{S_C}$ for all $(\alpha_m, \alpha_n)$.
We note that if the whole of a curve $\mathcal{C}$ (not necessarily closed) 
lies in a region of the no-pumping condition, 
then the average excess quantity does not depend on the whole of $\mathcal{C}$ 
but depends on only the initial and final points of $\mathcal{C}$.

\section{Application to spinless electron transport in quantum dot}\label{sec:Example}

In this section, we apply the general framework obtained in the previous section to 
the transport of spinless electrons in systems of coupled quantum dots with single-levels  
which are connected to two electron reservoirs ($b={\rm L}, {\rm R}$). 
The assumption of spinless electrons can be employed at least in the following two cases. 
One is that the intra-dot electron-electron interaction is quite large, 
so that each dot is at most singly occupied. 
In this case, if the system S and the reservoirs are spin-rotationally invariant, 
the spin degrees of freedom are irrelevant.
The other case is that a strong magnetic field is applied to the dots, 
so that the Zeeman energy of the electron spin is sufficiently large. 
In this case, since only the lower energy spin state for each dot is at most occupied 
and only the lower energy states are involved in the dynamics, 
the spin degrees of freedom is effectively negligible.

We assume that the reservoir Hamiltonian is given by 
$\hat{H}_b = \sum_k \hbar\Omega_{bk} \hat{c}_{bk}^\dag \hat{c}_{bk}$. 
Here $\hbar\Omega_{bk}$ is the $k$th mode energy of the electron in the $b$th reservoir,
and $\hat{c}_{bk}^\dag$  ($\hat{c}_{bk}$) is 
the corresponding creation (annihilation) operator, 
which satisfies $\{\hat{c}_{bk}^\dag, \hat{c}_{b'k'}\} = \delta_{kk'}\delta_{bb'}$ and 
$\{\hat{c}_{bk}^\dag, \hat{c}_{b'k'}^\dag\} = \{\hat{c}_{bk}, \hat{c}_{b'k'}\} = 0$. 
We here take the quantity to be counted as 
the electron number transferred from the reservoir L to the system S, 
i.e., $\hat{Q} = \hat{N}_{\rm L} = \sum_{k} \hat{c}_{Lk}^\dag \hat{c}_{Lk}$.

\subsection{Non-interacting electron model with RWA}

First we consider a model of non-interacting spinless electrons in a series of $N$ single-level quantum dots 
coupled to two reservoirs ($b={\rm L}, {\rm R}$). 
The Hamiltonian of the system S in this model is 
\begin{align}
\hat{H}_{\rm S} = \sum_i \varepsilon_i \hat{d}_i^\dag \hat{d}_i 
+ \sum_{ii'} ( v_{ii'} \hat{d}_i^\dag \hat{d}_{i'} + {\rm h.c.} ), 
\label{Hamiltonian_noint}
\end{align}
where $\varepsilon_i$ is the level energy of the $i$th dot, 
$\hat{d}_i^\dag$ ($\hat{d}_i$) is the creation (annihilation) operator of the electron in the $i$th dot, 
and $v_{ii'}$ is the transfer probability amplitude between the $i$th and $i'$th dots.
Note that we can use this Hamiltonian also for many-level quantum dots (with no electron-electron interaction) 
if we add the label of the intra-dot level to the label $i$ of the dot site, 
since it is always possible to rearrange the pairs of the labels of the site and level to be a single label.
Therefore we can also apply the results in this subsection to the non-interacting electron system 
in a series of the many-level quantum dots.

This Hamiltonian can be diagonalized by a unitary transformation as 
$\hat{H}_{\rm S} = \sum_j \hbar\omega_j \hat{a}_j^\dag \hat{a}_j$. 
Here $\hbar\omega_j$ is the $j$th mode energy of the electron in the system S, 
and $\hat{a}_j^\dag$ ($\hat{a}_j$) is the corresponding creation (annihilation) operator of the electron, 
which satisfies the canonical anti-commutation relations of fermion: 
$\{\hat{a}_{j}^\dag, \hat{a}_{j'}\} = \delta_{jj'}$ and 
$\{\hat{a}_{j}^\dag, \hat{a}_{j'}^\dag\} = \{\hat{a}_{j}, \hat{a}_{j'}\} = 0$. 
The coupling Hamiltonian between the system S and the $b$th reservoir is given by 
$\hat{H}_{{\rm S}b} = \sum_{j,k} V_{bjk} \hat{a}_j^\dag \hat{c}_{bk} + {\rm h.c.}$ 

To derive the GQME for this system we use the Born (up to the second order in $\hat{H}_{{\rm S}b}$)
and the Markov approximations \cite{BreuerPetruccione}.
Furthermore, we adopt the RWA, which is a coarse-graining of the time evolution 
on the time scale longer than that of the system evolution without the coupling with the reservoirs 
\cite{BreuerPetruccione,EspositoHarbolaMukamel}. 
For $\chi = 0$ the RWA leads a Lindblad form of the QME 
and guarantees the complete positivity of the time evolution.
We carry out the RWA by averaging over the rapidly oscillating terms in the Born-Markov GQME 
in the interaction picture (see Appendix~\ref{derivaton_GQME} for the detail). 
Then we obtain the generator $\mathcal{K}_\chi$ of the GQME 
in the form of $\mathcal{K}_{\chi} = \sum_j \mathcal{K}_{\chi,j}$. 
Here the GQME generator $\mathcal{K}_{\chi,j}$ for the $j$th mode is given by 
\cite{EspositoHarbolaMukamel}
\begin{align}
\mathcal{K}_{\chi,j} \hat{\rho} 
&\equiv \frac{1}{i\hbar} \bigl[ 
\hbar\omega_j \hat{a}_j^\dag \hat{a}_j + \hat{H}^{\rm Lamb}_j , \hat{\rho} \bigr]
\notag\\
- \frac{1}{2\hbar^2} & \sum_{b={\rm L}, {\rm R}} \Bigl( 
\Phi_{b j}^- (\omega_j) \bigl\{ 
\hat{a}_j^\dag \hat{a}_j \hat{\rho} + \hat{\rho} \hat{a}_j^\dag \hat{a}_j 
- 2 e^{-i\chi_b} \hat{a}_j \hat{\rho} \hat{a}_{j}^\dag \bigr\} 
\notag\\ 
&+ \Phi_{b j}^+ (\omega_j) \bigl\{ 
\hat{a}_j \hat{a}_j^\dag \hat{\rho} + \hat{\rho} \hat{a}_j \hat{a}_j^\dag 
- 2 e^{i\chi_b} \hat{a}_j^\dag \hat{\rho} \hat{a}_j 
\bigr\} \Bigr), 
\label{K_GQME_j}
\end{align}
where 
$\hat{H}^{\rm Lamb}_j \equiv \sum_b \bigl\{ \Psi_{b j}^- (\omega_j) \hat{a}_j^\dag \hat{a}_j 
- \Psi_{b j}^+ (\omega_j) \hat{a}_j \hat{a}_j^\dag \bigr\} / 2\hbar$ 
is the Lamb shift Hamiltonian, 
$\Phi_{b j}^\pm (\omega) \equiv 2 \pi \sum_k |V_{b jk}|^2 \delta (\omega - \Omega_{b k}) f_b^\pm(\omega)$ 
is the power spectrum of the $b$th reservoir, 
$\Psi_{b j}^\pm (\omega) \equiv P \int_{-\infty}^\infty (d\omega'/\pi) 
\Phi_{b j}^\pm ( \omega') / (\omega' - \omega)$, 
and $\chi_{\rm L} = \chi$, $\chi_{\rm R} = 0$.
Here, $P$ means the principal value, 
$f_b^+(\omega) = 1/(1+e^{-\beta_b(\hbar\omega -\mu_b)})$ 
is the fermi distribution function with $\beta_b , \mu_b$, 
and $f_b^-(\omega) = 1 - f_b^+(\omega)$.
The control parameters $\bm{\alpha}$ in this model can be both the system parameters, i.e., 
the levels $\{\varepsilon_i\}$ of the dots and the transfer $\{v_{ii'}\}$ between the dots, 
and the reservoir parameters $\{\beta_b, \mu_b\}$.

In this model within the RWA, we analytically obtain the BSN vector potential 
(see Appendix \ref{derivation_BSN_RWA} for the derivation): 
\begin{align}
\bm{A}(\bm{\alpha})
&= \sum_j \frac{\Gamma_{{\rm L}j}(\omega_j)}{\Gamma_j(\omega_j)} 
\frac{\partial}{\partial \bm{\alpha}} 
\left( \frac{\sum_b \Gamma_{b j}(\omega_j) f^+_b(\omega_j)}{\Gamma_j(\omega_j)} \right), 
\label{vectorpotentialRWA}
\end{align}
where $\Gamma_j(\omega_j) \equiv \sum_b \Gamma_{b j}(\omega_j)$, 
and $\Gamma_{b j}(\omega_j) \equiv \Phi_{b j}^+(\omega_j) + \Phi_{b j}^-(\omega_j)$ 
is the spectral function of the $b$th reservoir. 
We can also calculate the BSN curvature as 
\begin{align}
&F_{\alpha_m \alpha_n}
\notag\\
&= \sum_j \frac{\partial}{\partial \alpha_m} 
\left(\frac{\Gamma_{{\rm L}j}(\omega_j)}{\Gamma_j(\omega_j)} \right)
\frac{\partial}{\partial \alpha_n} 
\left( \frac{\sum_b \Gamma_{b j}(\omega_j) f^+_b(\omega_j)}{\Gamma_j(\omega_j)} \right).
\label{curvatureRWA}
\end{align}
From Eq.~(\ref{curvatureRWA}), 
we find that no net excess number of electrons flow per cycle if only the reservoir parameters 
$(\beta_{\rm L}, \mu_{\rm L}, \beta_{\rm R}, \mu_{\rm R})$ are modulated 
with the system parameters fixed. 
This is because the spectral function is written as 
$\Gamma_{b j}(\omega_j)= 2 \pi \sum_k |V_{b jk}|^2 \delta (\omega_j - \Omega_{b k})$, 
so that $\Gamma_{{\rm L}j} / \Gamma_j$ in Eq.~(\ref{curvatureRWA}) is independent of 
$(\beta_{\rm L}, \mu_{\rm L}, \beta_{\rm R}, \mu_{\rm R})$, 
and that Eq.~(\ref{vectorpotentialRWA}) can be written as the gradient of a scalar function 
of $(\beta_{\rm L}, \mu_{\rm L}, \beta_{\rm R}, \mu_{\rm R})$. 
We note that this result is characteristic of fermion systems. 
Indeed, in a model of single two-level system connected to two bosonic heat reservoirs, 
there exists heat pumping by cyclic modulations of the temperatures of the two reservoirs \cite{RenHanggiLi}.
This difference between the results for the fermion and boson reservoirs 
comes from the particle statistics, which leads that $\langle c^\dag c + c c^\dag \rangle_b$ 
is independent of (dependent on) the reservoir parameters for fermion (boson) reservoir.
Because $\Phi_{b j}^+$ and $\Phi_{b j}^-$ are respectively proportional to 
$\langle c^\dag c \rangle_b$ and $\langle c c^\dag \rangle_b$, 
the particle statistics determines whether $\Gamma_{b j}= \Phi_{b j}^+ + \Phi_{b j}^-$ 
depends on the reservoir parameters.

Before closing this subsection, we make a remark on the validity of the RWA on the transport systems.
The GQME gives the same result either with or without the RWA, 
as far as the transport between the system and the reservoirs is studied, 
whereas it is known that the internal current in the system vanishes 
in nonequilibrium steady states under the RWA \cite{Wichterich_etal}.
In Appendix~\ref{equivalence_RWA_NonRWA}, we analytically show that 
the unit-time cumulant generating function $\lambda^\chi_0(\bm{\alpha})$ 
of the quantity transferred from the reservoirs to the system 
in the steady state for fixed $\bm{\alpha}$ is equivalent between the GQMEs 
within and without the RWA. 
We also confirm numerically that the results of the adiabatic pumping within the RWA 
quantitatively agree with those without the RWA in the next subsection.

\subsection{Non-interacting electron in double quantum dot without RWA}

\begin{figure}[bt]
\begin{center}
\includegraphics[width=\linewidth]{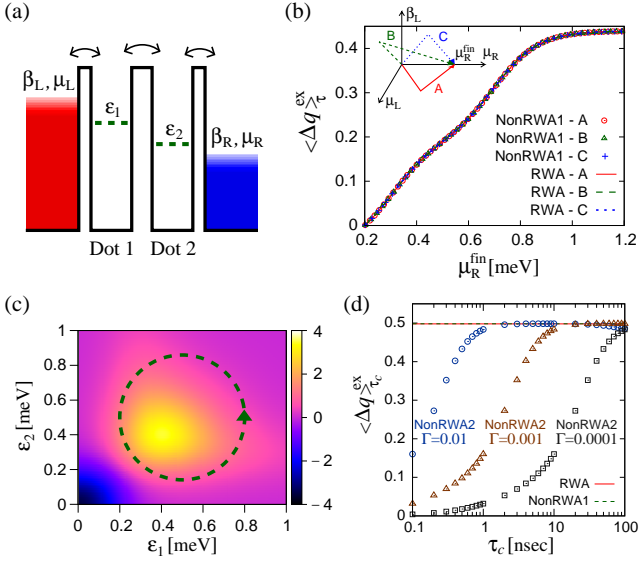}
\caption{(Color online) (a) Quantum double dot coupled to reservoirs. 
(b) Excess electron transfer $\langle \varDelta q \rangle_{\tau}^{\rm ex}$ 
when the reservoir parameters $\beta_{\rm L}$, 
$\mu_{\rm L}$, and $\mu_{\rm R}$ are slowly varied along the curves 
(A, B, and C) shown in the inset, where $\beta_{\rm R}$ and the system parameters 
are fixed ($\beta_{\rm R} = 1~[1/{\rm K}]$, 
$\varepsilon_1 = \varepsilon_2 = 0.5~[{\rm meV}]$, and 
$v = 0.2~[{\rm meV}]$). 
The initial values of the parameters are 
$\beta_{\rm L}^{\rm ini} = 1~[1/{\rm K}]$ and 
$\mu_{\rm L}^{\rm ini} = \mu_{\rm R}^{\rm ini} = 0.2~[{\rm meV}]$. 
The dependence on the final value $\mu_{\rm R}^{\rm fin}$ is plotted, 
where $\beta_{\rm L}^{\rm fin} = \beta_{\rm L}^{\rm ini}$ and 
$\mu_{\rm L}^{\rm fin} = \mu_{\rm L}^{\rm ini}$. 
The calculations are performed by RWA and NonRWA1.
(c) BSN curvature $F_{\varepsilon_1 \varepsilon_2}$ as a function of $\varepsilon_1$ 
and $\varepsilon_2$, calculated by NonRWA1. 
The reservoir parameters are set to 
$\beta_{\rm L} = \beta_{\rm R} = 1~[1/{\rm K}]$ and 
$\mu_{\rm L} = \mu_{\rm R} = 0.2~[{\rm meV}]$.
(d) Excess electron transfer $\langle \varDelta q \rangle_{\tau_c}^{\rm ex}$ 
for a cyclic process along the circle depicted in (c).
The horizontal axis is the period $\tau_c$ of the cyclic process.
The calculations are performed by RWA, NonRWA1, and NonRWA2.
The results by NonRWA2 are plotted for various values of the amplitude $\Gamma$
of the reservoir spectral function, 
while it is fixed to $0.001~[{\rm meV}]$ for the other methods.
}
\label{fig:DQD}
\end{center}
\end{figure}

Next we consider a non-interacting double quantum dot system coupled to two reservoirs, 
as illustrated in Fig.~\ref{fig:DQD}~(a) 
(the transfer probability amplitude between the dot 1 and 2 is denoted by $v$).
Here we assume the wide band limit, $\Gamma_{bj}(\omega) = \Gamma_b = {\rm const}.$ ($b=$L,R), 
and the symmetric coupling, $\Gamma_{\rm L} = \Gamma_{\rm R}=\Gamma$. 
In this subsection we use three different methods for calculating $\langle \varDelta q \rangle^{\rm ex}_\tau$ 
under modulations of the control parameters of the model.
In the first method (denoted by RWA), we apply our results within the RWA 
[Eqs.~(\ref{vectorpotentialRWA}) and (\ref{curvatureRWA})].
In the second one (denoted by NonRWA1), 
we numerically solve the eigenvalue problem of the GQME generator $\mathcal{K}_\chi$ 
without RWA, and use the geometrical formula (\ref{average_ex}).
In the third method (denoted by NonRWA2), we numerically solve the time evolution differential equation 
of the GQME without RWA, and use $S_\tau(\chi) = {\rm Tr}_{\rm S} \hat{\rho}^\chi(\tau)$.

In Fig.~\ref{fig:DQD}~(b), we plot the excess electron number $\langle \varDelta q \rangle^{\rm ex}_\tau$ 
transferred from the reservoir L to the system 
for non-cyclic modulations of $\beta_{\rm L}, \mu_{\rm L}$, and $\mu_{\rm R}$
along the curves illustrated in the inset of Fig.~\ref{fig:DQD}~(b).
We plot the dependence of $\langle \varDelta q \rangle^{\rm ex}_\tau$ 
on the final value of the right reservoir chemical potential $\mu_{\rm R}^{\rm fin}$ of the modulations.
We see that all the results agree within the numerical precision. 
This implies that the no-pumping condition described below Eq.~(\ref{curvatureRWA}) 
within the RWA still holds without the RWA. 
We have also confirmed that the absolute value of the BSN curvature 
computed without the RWA is less than $10^{-6}$ in the space 
of the reservoir parameters, which is zero within the numerical precision.

In Fig.~\ref{fig:DQD}~(c), we plot the BSN curvature $F_{\varepsilon_1 \varepsilon_2}$ 
calculated by the method of NonRWA1 as a function of $\varepsilon_1$ and $\varepsilon_2$.
We see that the curvature takes the non-zero values in this case. 
Therefore we have a finite geometrical pumping for the slow periodic modulation 
of the dot levels (system parameters).
We note that this result of the BSN curvature also agrees with the RWA result 
given by Eq.~(\ref{curvatureRWA}), although not shown in the figure. 

We also calculate $\langle \varDelta q \rangle^{\rm ex}_{\tau_c}$ for the cyclic process 
depicted in Fig.~\ref{fig:DQD}~(c).
In Fig.~\ref{fig:DQD}~(d), we plot $\langle \varDelta q \rangle^{\rm ex}_{\tau_c}$ 
calculated by NonRWA2 as a function of the cycle period $\tau_c$ 
for various values of the amplitude $\Gamma$ of the spectral function.
We see that for large $\tau_c$ the asymptotic results by NonRWA2 agree with the results by RWA and NonRWA1.
This supports the validity of the adiabatic approximation 
used in deriving Eq.~(\ref{average_ex}) for slow modulations.
We also see that the characteristic time for the validity of the adiabatic approximation 
becomes shorter as $\Gamma$ increases. 
This implies that $\tau_c \gg \tau_{\rm rlx} \equiv \hbar/\Gamma$ is the adiabatic condition, 
as is mentioned below Eq.~(\ref{eq_c0}) in the previous section.

We note that all the results by RWA and NonRWA1 agree with each other 
not only qualitatively but also quantitatively. 
This implies that the rotating wave approximation is valid in discussing transport 
between the system and the reservoirs under slow modulations of the parameters.

\subsection{Interacting electron model}\label{sec:interacting_model}

\begin{figure}[bt]
\begin{center}
\includegraphics[width=\linewidth]{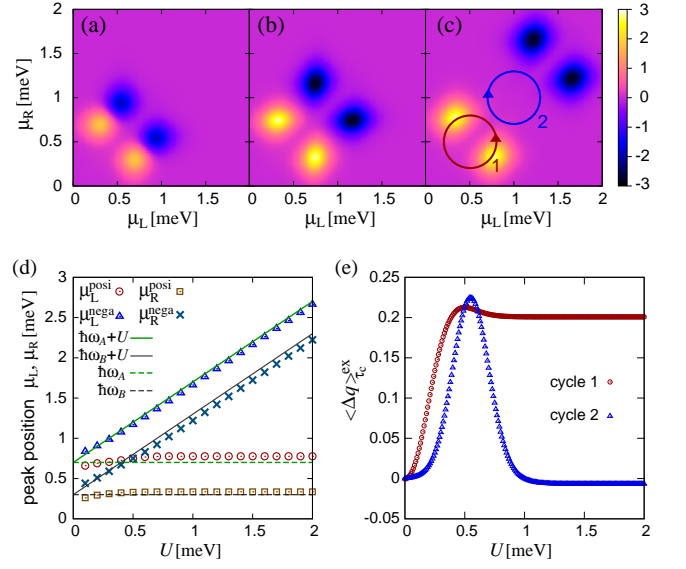}
\caption{(Color online) 
(a)-(c) BSN curvature $F_{\mu_{\rm L} \mu_{\rm R}}$ for the interacting electron system 
in the double dot, as a function of $\mu_{\rm L}$ and $\mu_{\rm R}$.
The interaction strength is $U=0.25[{\rm meV}]$ for (a), 
$U=0.5[{\rm meV}]$ for (b), and $U=1.0[{\rm meV}]$ for (c). 
The other parameters are $\beta_{\rm L} = \beta_{\rm R} = 1~[1/{\rm K}]$, 
$\varepsilon_1 = \varepsilon_2 = 0.5~[{\rm meV}]$, $v = 0.2~[{\rm meV}]$,  
and $\Gamma = 0.001~[{\rm meV}]$. 
(d) Positions of the peaks of the BSN curvature, as a function of $U$.  
The position ($\mu_{\rm L}^{\rm posi}$, $\mu_{\rm R}^{\rm posi}$) of 
the positive peak with $\mu_{\rm L}^{\rm posi} > \mu_{\rm R}^{\rm posi}$
and that ($\mu_{\rm L}^{\rm nega}$, $\mu_{\rm R}^{\rm nega}$) 
of the negative peaks with $\mu_{\rm L}^{\rm nega} > \mu_{\rm R}^{\rm nega}$ are shown.
The solid lines are $\hbar\omega_A+U$ and $\hbar\omega_B+U$, 
and the dashed lines are the mode energies, $\hbar\omega_A$ and $\hbar\omega_B$, 
of the system Hamiltonian ($\hbar\omega_A > \hbar\omega_B$). 
(e) Excess electron transfer $\langle \varDelta q \rangle_{\tau_c}^{\rm ex}$ 
for cyclic processes along the circles (1 and 2) depicted in (c).
The horizontal axis is the interaction strength $U$. 
The direction of the cycle~2 is opposite to that of the cycle~1.
}
\label{fig:DQDInteraction}
\end{center}
\end{figure}

We next consider an interacting spinless electron system in a double quantum dot.
As is mentioned at the beginning of this section, 
we can use the single-level dot model under certain conditions. 
Even in this case, the inter-dot interaction exists, 
and in some situations (e.g., short inter-dot distance) it is not negligible.
We here consider such a situation, where the system Hamiltonian is given by 
\begin{align}
\hat{H}_{\rm S} = \sum_{i=1,2} \varepsilon_i \hat{d}_i^\dag \hat{d}_i 
+ v (\hat{d}_1^\dag \hat{d}_2 + \hat{d}_2^\dag \hat{d}_1) 
+ U \hat{d}_1^\dag \hat{d}_1 \hat{d}_2^\dag \hat{d}_2. 
\end{align}
In this model, an electron in one dot interacts with an electron in the other dot.
We investigate the excess electron transfer 
$\langle \varDelta q \rangle_{\tau_c}^{\rm ex}$ 
under the modulation of the chemical potentials ($\mu_{\rm L}$ and $\mu_{\rm R}$).
We also assume the wide band limit and symmetric coupling: $\Gamma_{bj}(\omega) = \Gamma$ ($b=$L,R). 
This model is essentially the same as that considered in Sec.~III of Ref.~\onlinecite{RiwarSplettstoesser}.
In Appendix~\ref{sec:Riwar}, we will check the consistency of the results calculated in our scheme 
with the results in Ref.~\onlinecite{RiwarSplettstoesser}.

In Fig.~\ref{fig:DQDInteraction}~(a)-(c), we plot the BSN curvature $F_{\mu_{\rm L} \mu_{\rm R}}$ 
for various values of the interaction strength $U$ 
obtained from the numerical diagonalization method without the RWA.
We find that there exist two positive and two negative peaks. 
In Fig.~\ref{fig:DQDInteraction}~(d), we plot the position of 
one of the positive peaks and that of one of the negative peaks.
We see that the positions of the negative peaks move as $U$ increases, 
whereas those of the positive ones do not.
Moreover we find that these peak positions are located around at the energies necessary to add one electron; 
the positions of the positive peaks are 
$(\mu_{\rm L},\mu_{\rm R}) = (\hbar\omega_A,\hbar\omega_B)$ and  $(\hbar\omega_B,\hbar\omega_A)$, 
and those of the negative ones are $(\mu_{\rm L},\mu_{\rm R}) = (\hbar\omega_A+U,\hbar\omega_B+U)$ 
and $(\hbar\omega_B+U,\hbar\omega_A+U)$, 
where the mode energies are given by
$\hbar\omega_{A,B} 
= (1/2) \bigl\{ \varepsilon_1 + \varepsilon_2 \pm \sqrt{(\varepsilon_1 - \varepsilon_2)^2 + 4v^2} \bigr\}$ 
(the subscripts $A$ and $B$ stand for the anti-bonding and bonding modes, respectively).
This implies that the positive and negative peaks merge for non-interacting system ($U=0$), 
and thus the curvature $F_{\mu_{\rm L} \mu_{\rm R}}$ becomes zero in $(\mu_{\rm L},\mu_{\rm R})$-space 
to achieve the no-pumping condition mentioned below Eq.~(\ref{curvatureRWA}).
This result also implies that, for $U>0$, an adiabatic pumping can occur 
even if only the reservoir parameters are modulated.

Indeed, Fig.~\ref{fig:DQDInteraction}~(e) shows the $U$-dependence of 
the excess electron transfer $\langle \varDelta q \rangle_{\tau_c}^{\rm ex}$ 
for cyclic processes [cycle~1 and 2 depicted in Fig.~\ref{fig:DQDInteraction}~(c)], 
where $\langle \varDelta q \rangle_{\tau_c}^{\rm ex}$ is non-zero for $U>0$. 
Note that the direction of the cycle~2 is opposite to that of the cycle~1.
We observe that $\langle \varDelta q \rangle_{\tau_c}^{\rm ex}$ for the cycle~1 
becomes a constant for $U \gtrsim 1[{\rm meV}]$. 
In contrast, $\langle \varDelta q \rangle_{\tau_c}^{\rm ex}$ for the cycle~2 has 
a peak at $U \simeq 0.5[{\rm meV}]$ and becomes nearly zero for $U \gtrsim 1[{\rm meV}]$. 
These results are consistent with the behaviors of the peak positions 
of the BSN curvature $F_{\mu_{\rm L} \mu_{\rm R}}$ 
shown in Fig.~\ref{fig:DQDInteraction}~(a)-(c) and (d): 
as $U$ increases, 
the positive peak positions of $F_{\mu_{\rm L} \mu_{\rm R}}$ stay around the location of the cycle~1 
whereas the negative peak positions of $F_{\mu_{\rm L} \mu_{\rm R}}$ pass across the location of the cycle~2.

Qualitatively, these results are understood as follows.
The pumped current resonantly flows when the chemical potential of the reservoirs
agrees with the energy necessary to add (or remove) one electron \cite{Kashcheyevs_etal}. 
In the present model, these energies are 
$\hbar\omega_B$ for the transition $|0\rangle \leftrightarrow |B\rangle$, 
$\hbar\omega_A$ for $|0\rangle \leftrightarrow |A\rangle$, 
$\hbar\omega_B+U$ for $|A\rangle \leftrightarrow |D\rangle$, 
and $\hbar\omega_A+U$ for $|B\rangle \leftrightarrow |D\rangle$. 
Here $|0\rangle$, $|A\rangle$, $|B\rangle$, and $|D\rangle$ are the eigenstates of $\hat{H}_{\rm S}$ 
(the empty, anti-bonding, bonding, and doubly occupied states, respectively).
The transitions $|0\rangle \leftrightarrow |B\rangle$ and $|A\rangle \leftrightarrow |D\rangle$ 
(or $|0\rangle \leftrightarrow |A\rangle$ and $|B\rangle \leftrightarrow |D\rangle$) 
have opossitely directed contribution to the pumping.
For $U=0$, since these two resonant points locate at the same position, 
the contribution from these two cancels out.
The inter-dot interaction breaks this degeneracy of the resonant points; 
for $U \neq 0$, the locations of the resonant points separate and thus non-zero pumped current can flow.
The non-monotonic behavior for the cycle~2 in Fig.~\ref{fig:DQDInteraction}~(e) 
can be also understood as follows. 
In the smaller $U$ region, the resonant points come into the cycle~2 as $U$ increases, 
which results in the increase of $\langle \varDelta q \rangle_{\tau_c}^{\rm ex}$ in this region. 
In the larger $U$ region, on the other hand, the resonant points go out of the cycle~2 as $U$ increases, 
which results in the decrease of $\langle \varDelta q \rangle_{\tau_c}^{\rm ex}$ in this region.

\section{Discussion and Conclusion}\label{sec:Conclusion}

By using a QME approach, we have derived a geometrical expressions of the cumulant generating function 
and average of the pumped (excess) quantity transferred from reservoirs to the system 
under slow modulation of control parameters, 
where the BSN phases, vector potential, and the curvature of the QME play crucial roles.

For non-interacting electrons in quantum dot systems, there is no pumped current 
when only the temperatures and chemical potentials of the reservoirs are modulated. 
In contrast, for an interacting system, the pumped current can be observed even in this situation.
We note that the modulations of only the chemical potentials of the reservoirs are required for the pumping 
(in the interacting system). 
This has an advantage for the control of the pumping in experiments, 
since the modulation of chemical potential is easier than that of the temperatures.
As shown in Fig.~\ref{fig:DQDInteraction}~(e), for a cyclic modulation of the chemical potentials, 
the pumped current depends not only on the difference of the chemical potentials but also on their average. 
This implies that the average number of electrons in the system S is important for the pumping.
For example, when we modulate the chemical potentials as 
$\mu_{\rm L}(t) = \mu_c + \mu_{\rm rad}\cos(2\pi t /\tau_c)$ and 
$\mu_{\rm R}(t) = \mu_c + \mu_{\rm rad}\sin(2\pi t /\tau_c)$, 
$\mu_c$ affects the quantity of the pumping.
This fact may be applicable for switching the pumping by the change of $\mu_c$ or 
the electron density, which can be controlled by a gate voltage.

Since we have employed the method of the full counting statistics, 
we can also calculate the fluctuation (noise) of the pumped quantity.
It is a future issue to analyse the detailed properties of the fluctuation in the adiabatic pumping.
Although we have applied our formulation only to the examples of spinless systems in this paper, 
we can apply it also to the QME description of spin pumping 
\cite{Watson_etal,Cota_etal,RiwarSplettstoesser,Splettstoesser2008,Deus_etal}.  
It is also interesting to investigate the relation 
between the present geometrical expression of the adiabatic pumping based on the QME 
and the conventional geometrical expressions based on the scattering theory 
\cite{Brouwer,Avron_etal,AndreevKamenev,MakhlinMirlin}, 
and to clarify the condition for quantized charge pumping (topological effect) 
as in the case of the classical master equation \cite{Chernyak_etal1,Chernyak_etal2}.
The investigations of non-adiabatic pumping, non-Markovian situation, and spin effect are also future issues.
For example, it is important to consider the electron system with spin 
in a single dot with the on-site Hubbard Hamiltonian (Anderson model) 
\cite{Aono,Splettstoesser2005,Splettstoesser2006,Reckermann_etal}, 
and to compare with our results of spinless case \cite{Yoshii}.

\begin{acknowledgments}
The authors acknowledge K. Saito, R. Yoshii, and M. Yamaguchi for their helpful advice. 
This work was supported by 
the JSPS Research Fellowships for Young Scientist (No. 24-1112) and 
the Grant-in-Aid for Research Activity Start-up (KAKENHI 11025807). 
\end{acknowledgments}

\appendix

\section{Generalized quantum master equation without and within the RWA}
\label{derivaton_GQME}

We here derive the concrete form of the GQME (\ref{GQME}) and (\ref{GQMEgenerator}) without and within the RWA.

We start from the total Hamiltonian (system plus reservoirs) given as 
\begin{align}
\hat{H}_{\rm tot} &= \hat{H}_{\rm S} + \sum_{b} \hat{H}_b 
+ u \sum_{b} \hat{H}_{{\rm S}b}.
\end{align}
Here, for simplicity, we assume that the reservoir Hamiltonian $\hat{H}_b$ 
and the coupling Hamiltonian $\hat{H}_{{\rm S}b}$ between the system and the reservoir 
are respectively written as 
\begin{align}
\hat{H}_b &= \sum_k \hbar\Omega_{bk} \hat{c}_{bk}^\dag \hat{c}_{bk}, 
\\
\hat{H}_{{\rm S}b} &= \sum_k \bigl( V_{bk} \hat{a}_{i_b}^\dag \hat{c}_{bk} 
+ V_{bk}^* \hat{a}_{i_b} \hat{c}_{bk}^\dag \bigr), 
\end{align}
where $\hat{a}_i$ is a single-particle mode annihilation operator in the system S, 
$i_b$ is the index of the system mode that couples to the $b$th reservoir, 
and $\hat{c}_{bk}$ is the $k$th mode annihilation operator in the $b$th reservoir.
We denote the eigenenergy of the system Hamiltonian $\hat{H}_{\rm S}$ as $E_x$, 
and the corresponding energy eigenstate as $|E_x\rangle$.
We also assume that all the eigenenergies of $\hat{H}_{\rm S}$ are non-degenerate. 
We consider the quantity 
$\hat{Q} = \sum_b\sum_k q_{bk}\hat{c}_{bk}^\dag \hat{c}_{bk}$ 
to be counted, and define the current of $\hat{Q}$ from the reservoirs to the system S as positive.

For the derivation of the GQME, 
it is convenient to introduce the eigenoperators \cite{BreuerPetruccione} from $\hat{a}_{i_b}$:
\begin{align}
\hat{a}_{i_b}^{(\omega_{\rm S})} &= \sum_{E_x} |E_x - \hbar\omega_{\rm S} \rangle 
\langle E_x - \hbar\omega_{\rm S} | \hat{a}_{i_b} |E_x \rangle \langle E_x | , 
\\
\hat{a}_{i_b}^{\dag (\omega_{\rm S})} &= \sum_{E_x} |E_x + \hbar\omega_{\rm S} \rangle 
\langle E_x + \hbar\omega_{\rm S} | \hat{a}_{i_b}^\dag |E_x \rangle \langle E_x |.
\end{align}
Then the modified coupling Hamiltonian in the interaction picture is written as
\begin{align}
\Check{H}_{{\rm S}b}^\chi (t) &= 
e^{-i \chi \hat{Q} /2} e^{-(\hat{H}_{\rm S} + \hat{H}_b) t / {\rm i}\hbar} 
\hat{H}_{{\rm S}b}^\chi e^{(\hat{H}_{\rm S} + \hat{H}_b) t / {\rm i}\hbar}
e^{i \chi \hat{Q} /2} 
\notag\\
&= \sum_k \sum_{\omega_{\rm S}} 
\Bigl( V_{bk} \hat{a}_{i_b}^{\dag(\omega_{\rm S})} \hat{c}_{bk} 
e^{i \chi q_{bk}/2} e^{i(\omega_{\rm S} - \Omega_{bk})t}
\notag\\
&+ V_{bk}^* \hat{a}_{i_b}^{(\omega_{\rm S})} \hat{c}_{bk}^\dag 
e^{-i \chi q_{bk}/2} e^{-i(\omega_{\rm S} - \Omega_{bk})t} \Bigr).
\label{H_Sb_chi}
\end{align}

We assume that the initial state of the total system is written 
as $\hat{\rho}_{\rm tot}(0) = \hat{\rho}_0 \otimes \hat{\rho}_{\rm res}$, 
where $\hat{\rho}_0$ is an initial state of the system S, 
$\hat{\rho}_{\rm res} = \bigotimes_b \hat{\rho}_b^{\rm G}$, and 
$\hat{\rho}_b^{\rm G} = e^{-\beta_b (\hat{H}_b - \mu_b \hat{N}_b) } / Z_b$
is the grand-canonical state of the $b$th reservoir.
Then substituting Eq.~(\ref{H_Sb_chi}) into Eq.~(\ref{GQME}), we obtain the GQME 
\begin{widetext}
\begin{align}
\frac{d}{dt} \hat{\rho}^\chi(t) 
&= \frac{1}{i\hbar} \bigl[ \hat{H}_{\rm S} , \hat{\rho}^\chi(t) \bigr]
\notag\\
- \frac{u^2}{2\hbar^2} & \sum_{b} \sum_{\omega_{\rm S}\omega_{\rm S}'} 
\Bigl[ \tilde{\Phi}_{b}^+ (\omega_{\rm S}') \Bigl\{ 
\hat{a}_{i_b}^{(\omega_{\rm S})} \hat{a}_{i_b}^{\dag(\omega_{\rm S}')} 
\hat{\rho}^\chi(t) 
+ \hat{\rho}^\chi(t) \hat{a}_{i_b}^{(\omega_{\rm S}')} 
\hat{a}_{i_b}^{\dag(\omega_{\rm S})}  
- e^{{\rm i} \chi q(\omega_{\rm S}')} \Bigl( 
\hat{a}_{i_b}^{\dag(\omega_{\rm S})} \hat{\rho}^\chi(t) 
\hat{a}_{i_b}^{(\omega_{\rm S}')} 
+ \hat{a}_{i_b}^{\dag(\omega_{\rm S}')} \hat{\rho}^\chi(t) 
\hat{a}_{i_b}^{(\omega_{\rm S})} 
\Bigr) \Bigr\}
\notag\\
&
+ \tilde{\Phi}_{b}^- (\omega_{\rm S}') \Bigl\{ 
\hat{a}_{i_b}^{\dag(\omega_{\rm S})} \hat{a}_{i_b}^{(\omega_{\rm S}')} 
\hat{\rho}^\chi(t) 
+ \hat{\rho}^\chi(t) \hat{a}_{i_b}^{\dag(\omega_{\rm S}')} 
\hat{a}_{i_b}^{(\omega_{\rm S})}  
- e^{-{\rm i} \chi q(\omega_{\rm S}')} \Bigl( 
\hat{a}_{i_b}^{(\omega_{\rm S})} \hat{\rho}^\chi(t) 
\hat{a}_{i_b}^{\dag(\omega_{\rm S}')} 
+ \hat{a}_{i_b}^{(\omega_{\rm S}')} \hat{\rho}^\chi(t) 
\hat{a}_{i_b}^{\dag(\omega_{\rm S})} 
\Bigr) \Bigr\}
\notag\\
&
+ i\tilde{\Psi}_{b}^+ (\omega_{\rm S}') \Bigl\{ 
- \hat{a}_{i_b}^{(\omega_{\rm S})} \hat{a}_{i_b}^{\dag(\omega_{\rm S}')} 
\hat{\rho}^\chi(t) 
+ \hat{\rho}^\chi(t) \hat{a}_{i_b}^{(\omega_{\rm S}')} 
\hat{a}_{i_b}^{\dag(\omega_{\rm S})} 
\Bigr\} 
+ i\tilde{\Psi}_{b}^+ (\omega_{\rm S}';\chi) \Bigl\{ 
- \hat{a}_{i_b}^{(\omega_{\rm S})} \hat{\rho}^\chi(t) 
\hat{a}_{i_b}^{\dag(\omega_{\rm S}')} 
+ \hat{a}_{i_b}^{(\omega_{\rm S}')} \hat{\rho}^\chi(t) 
\hat{a}_{i_b}^{\dag(\omega_{\rm S})} 
\Bigr\} 
\notag\\
&
+ i\tilde{\Psi}_{b}^- (\omega_{\rm S}') \Bigl\{ 
- \hat{a}_{i_b}^{\dag(\omega_{\rm S})} \hat{a}_{i_b}^{(\omega_{\rm S}')} 
\hat{\rho}^\chi(t) 
+ \hat{\rho}^\chi(t) \hat{a}_{i_b}^{\dag(\omega_{\rm S}')} 
\hat{a}_{i_b}^{(\omega_{\rm S})} 
\Bigr\} 
+ i\tilde{\Psi}_{b}^- (\omega_{\rm S}';\chi) \Bigl\{ 
- \hat{a}_{i_b}^{\dag(\omega_{\rm S})} \hat{\rho}^\chi(t) 
\hat{a}_{i_b}^{(\omega_{\rm S}')} 
+ \hat{a}_{i_b}^{\dag(\omega_{\rm S}')} \hat{\rho}^\chi(t) 
\hat{a}_{i_b}^{(\omega_{\rm S})} 
\Bigr\} \Bigr],
\label{Appendix:GQME}
\end{align}
\end{widetext}
where we used 
\begin{align}
\int_{0}^\infty {\rm d}t' e^{{\rm i}\omega t'} 
= \pi\delta(\omega) + {\rm i}\frac{P}{\omega}.
\end{align}
Here, $q(\Omega_{bk}) \equiv q_{bk}$, and 
\begin{align}
\tilde{\Phi}_{b}^\pm (\omega) 
&= \sum_{k} 2\pi \delta(\Omega_{bk} - \omega) |V_{bk}|^2 f_{bk}^\pm , 
\\
\tilde{\Psi}_{b}^\pm (\omega) 
&
= 2 \sum_{k} P\frac{|V_{bk}|^2 f_{bk}^\pm}{\Omega_{bk} - \omega} , 
\end{align}
\begin{align}
\tilde{\Psi}_{b}^\pm (\omega;\chi) 
&
= 2 \sum_{k} P\frac{|V_{bk}|^2 f_{bk}^\pm}{\Omega_{bk} - \omega} 
e^{\pm i \chi q(\Omega_{bk})},
\end{align}
\begin{align}
f_{bk}^+ &={\rm Tr}_b \Bigl\{ \hat{\rho}_b 
\hat{c}_{bk}^\dag \hat{c}_{bk} \Bigr\} 
= \frac{1}{1+e^{\beta_b(\hbar\Omega_{bk}-\mu_b)}},
\\
f_{bk}^- &={\rm Tr}_b \Bigl\{ \hat{\rho}_b 
\hat{c}_{bk} \hat{c}_{bk}^\dag \Bigr\}  = 1 - f_{bk}^+ . 
\end{align}
Equation~(\ref{Appendix:GQME}) is the concrete form of the GQME without the RWA.

When we transform Eq.~(\ref{Appendix:GQME}) in the interaction picture, 
we see that rapidly oscillating terms proportional to $\exp[\pm i(\omega_{\rm S}-\omega_{\rm S}')t]$ appear. 
In the RWA we neglect these terms \cite{BreuerPetruccione}.
Thus we obtain the GQME with the RWA by leaving only the terms with $\omega_{\rm S}' = \omega_{\rm S}$
in Eq.~(\ref{Appendix:GQME}). 

For the non-interacting models in Secs.~\ref{sec:Example}~A and B, 
because the eigenoperators are the mode operators themselves, 
the GQMEs for these models, in particular Eq.~(\ref{K_GQME_j}) for the RWA case, are derived.

\section{Derivation of Eq.~(\ref{vectorpotentialRWA})}\label{derivation_BSN_RWA}

In the model in Sec.~\ref{sec:Example}~A within the RWA, 
the eigenvalues and the eigenvectors of $\mathcal{K}_{\chi}$ 
can be decomposed into those of $\mathcal{K}_{\chi,j}$. 
That is, $\lambda_0^\chi = \sum_j \lambda_{0,j}^\chi$, 
$\hat{\ell}_0^\chi = \bigotimes_j \hat{\ell}_{0,j}^\chi$, 
and $\hat{\rho}_0^\chi = \bigotimes_j \hat{\rho}_{0,j}^\chi$, 
where $\lambda_{0,j}^\chi$ is the eigenvalue of $\mathcal{K}_{\chi,j}$ with maximum real part, 
and $\hat{\ell}_{0,j}^\chi$ and $\hat{\rho}_{0,j}^\chi$ 
are respectively the corresponding left and right eigenvectors, 
which are operators on the $j$th mode Hilbert space.

When we represent the left and right eigenvectors in the basis of the number states 
(denoted by $|0_j\rangle$ and $|1_j\rangle$) of $\hat{a}_j$ 
such that $\hat{a}_j |0_j\rangle = 0$ and $|1_j\rangle = \hat{a}_j^\dag |0_j\rangle$, we can show that 
$\langle m_j | \hat{\ell}_{0,j}^\chi | (1-m)_j \rangle 
= \langle m_j | \hat{\rho}_{0,j}^\chi | (1-m)_j \rangle = 0$ 
($m=0,1$), and 
\begin{align}
\begin{pmatrix}
\langle 0_j | \hat{\ell}_{0,j}^\chi | 0_j \rangle 
\\
\langle 1_j | \hat{\ell}_{0,j}^\chi | 1_j \rangle 
\end{pmatrix}
&= 
\begin{pmatrix}
1 \\
v_j(\chi)
\end{pmatrix}, 
\\
\begin{pmatrix}
\langle 0_j | \hat{\rho}_{0,j}^\chi | 0_j \rangle 
\\
\langle 1_j | \hat{\rho}_{0,j}^\chi | 1_j \rangle 
\end{pmatrix}
&= C_j(\chi)
\begin{pmatrix}
1 \\
w_j(\chi)
\end{pmatrix}, 
\end{align}
where 
\begin{align*}
v_j(\chi) 
&\equiv \frac{\sum_b \bigl(\Phi_{b j}^+(\omega_j) - \Phi_{b j}^-(\omega_j)\bigr) 
+ \sqrt{D_j}}
{2 \bigl(\Phi_{{\rm L}j}^+(\omega_j) e^{i \chi} + \Phi_{{\rm R}j}^+(\omega_j)\bigr)} , 
\\
w_j(\chi) 
&\equiv \frac{\sum_b \bigl(\Phi_{b j}^+(\omega_j) - \Phi_{b j}^-(\omega_j)\bigr) 
+ \sqrt{D_j}}
{2 \bigl(\Phi_{{\rm L}j}^-(\omega_j) e^{-i \chi} + \Phi_{{\rm R}j}^-(\omega_j)\bigr)} , 
\\
D_j &\equiv \Gamma_j{}^2 
- 4 ( 1 - e^{i \chi}) \Phi_{{\rm L}j}^+(\omega_j) \Phi_{{\rm R}j}^-(\omega_j) 
\notag\\
&- 4 ( 1 - e^{-i \chi}) \Phi_{{\rm L}j}^-(\omega_j) \Phi_{{\rm R}j}^+(\omega_j) , 
\end{align*}
$\Gamma_j = \sum_b \Gamma_{b j}$, and $\Gamma_{b j} = \Phi_{b j}^+(\omega_j) + \Phi_{b j}^-(\omega_j)$. 
From the normalization condition for $\chi=0$, ${\rm Tr}_{\rm S} \hat{\rho}_{0,j}^0=1$, 
we have $C_j(0) = \sum_b \Phi_{b j}^-(\omega_j) / \Gamma_j$.
Thus we obtain the BSN vector potential: 
\begin{align}
\bm{A}(\bm{\alpha})
&= \sum_j \frac{\partial v_j(\chi)}{\partial(i\chi)}\bigg|_{\chi=0} 
\frac{\partial \bigl(C_j(0)w_j(0)\bigr)}{\partial\bm{\alpha}}, 
\end{align}
which becomes the desired result after straightforward calculation.

\section{Equivalence of the unit-time cumulant generating functions 
without and within the RWA}\label{equivalence_RWA_NonRWA}

\subsection{Matrix representation of GQME generator}

To show the equivalence, we introduce the matrix representation of $\mathcal{K}_\chi$ 
by using the eigenstates $|E_x\rangle$ of $\hat{H}_{\rm S}$: 
the $(y'y,x'x)$ matrix element is given as 
$(\mathcal{K}_\chi)_{y'y,x'x} \equiv {\rm Tr}_{\rm S} \Bigl[ \bigl(|E_{y'} \rangle \langle E_y|\bigr)^\dag 
\bigl(\mathcal{K}_\chi |E_{x'} \rangle \langle E_x|\bigr) \Bigr]$, 
where $xx'$ ($yy'$) is the index for the column (row) of the matrix.

In this representation, we can show that within the RWA 
$(\mathcal{K}_\chi^{\rm RWA})_{y'y,xx} = (\mathcal{K}_\chi^{\rm RWA})_{yy,x'x} = 0$
if $x'\neq x$ and $y'\neq y$.
This implies that $\mathcal{K}_\chi^{\rm RWA}$ is a block diagonal matrix 
that is composed of $\{(\mathcal{K}_\chi^{\rm RWA})_{yy,xx}\}$ and 
$\{(\mathcal{K}_\chi^{\rm RWA})_{y'y,x'x}\}$ with $x'\neq x$ and $y'\neq y$.
We also note that a relation holds between the matrices of the generators without and with the RWA: 
$(\mathcal{K}_\chi^{\rm RWA})_{yy,xx} = (\mathcal{K}_\chi)_{yy,xx}$.

\subsection{Equivalence of $\lambda_0^\chi$ without and within RWA}

As is mentioned in Sec.~\ref{sec:GeneralResults}~B, 
the unit-time cumulant generating function in a steady state is given by 
the eigenvalue $\lambda_0^\chi$ of the generator $\mathcal{K}_\chi$ with maximum real part.

Within the RWA, $\lambda_0^\chi$ is determined from 
the eigenvalues of $\{(\mathcal{K}_\chi^{\rm RWA})_{yy,xx}\}$, 
one of the blocks of $\{(\mathcal{K}_\chi^{\rm RWA})_{yy',xx'}\}$. 

Without the RWA, the eigenvalues of $\mathcal{K}_\chi$ is determined 
by a perturbation theory with respect to $\nu=u^2$.
From Eq.~(\ref{Appendix:GQME}), we see that the unperturbed part of $\mathcal{K}_\chi$ is 
$- \{(E_x - E_{x'}) / i\hbar\} \delta_{E_y,E_x} \delta_{E_{y'},E_{x'}}$, and is diagonal.
Therefore the unperturbed eigenvalue is $-(E_x - E_{x'})/i\hbar$. 
This implies that the eigenvalue of zero has $d$-fold degeneracy, 
where $d$ is the dimension of the Hilbert space of the system S. 
Thus by the perturbation theory for degenerate case, 
the first order eigenvalue is determined by the eigenvalue equation for the matrix in the degenerate subspace, 
i.e., $\{(\mathcal{K}_\chi)_{yy,xx}\}$.
Furthermore, since the relation $(\mathcal{K}_\chi)_{yy,xx} = (\mathcal{K}_\chi^{\rm RWA})_{yy,xx}$ holds, 
the first order eigenvalues are equivalent to those of $\{(\mathcal{K}_\chi^{\rm RWA})_{yy,xx}\}$.
Therefore $\lambda_0^\chi$ without the RWA is equivalent to $\lambda_0^\chi$ within the RWA in $O(\nu)$.
This verification of the equivalence is sufficient since the master equation is valid up to $O(\nu)$.

\section{Comparison with result in another scheme}\label{sec:Riwar}

\begin{figure}[t]
\begin{center}
\includegraphics[width=0.9\linewidth]{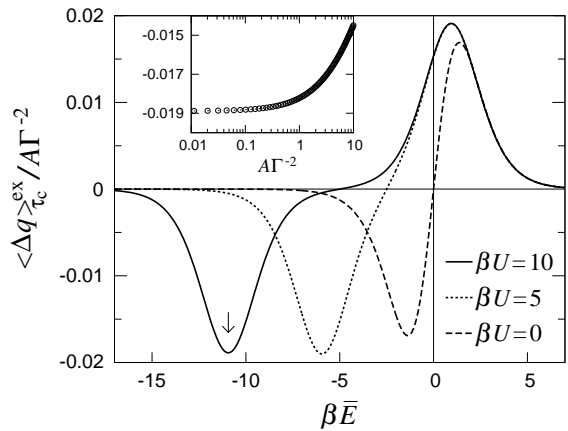}
\caption{
Electron transfer $\langle \Delta q \rangle^{\rm ex}_{\tau_c}$ normalized by $A \Gamma^{-2}$ 
per cycle of modulation of the dot levels, where $A$ is the cycle area in the parameter space 
($A\Gamma^{-2}=\pi\times 10^{-2}$ is used in the main plot).
The horizontal axis is the cycle center $\bar{E}$ multiplied by the inverse temperature $\beta$ of the reservoirs.
The results for various values of the inter-dot interaction $U$ are plotted.
The inset shows $\langle \Delta q \rangle^{\rm ex}_{\tau_c}/A \Gamma^{-2}$ at the negative peak 
for $\beta U=10$ (indicated by the arrow in the main) as a function of $A\Gamma^{-2}$.
The values of the parameters are as follows: $\beta_{\rm L} = \beta_{\rm R} = \beta = 50$~[1/meV], 
$\beta \mu_{\rm L} = \beta \mu_{\rm R} = 0$, $\beta v = 1/2$, and $\beta\Gamma = 1/2$. 
}
\label{fig:Riwar}
\end{center}
\end{figure}

Here we check the consistency of the results in our scheme with those in Ref.~\onlinecite{RiwarSplettstoesser}.
We again consider the model of interacting double quantum dot in Sec.~\ref{sec:interacting_model}. 
Under the condition of the symmetric reservoirs 
($\beta_{\rm L}=\beta_{\rm R}=\beta$, $\mu_{\rm L}=\mu_{\rm R}=0$, and $\Gamma_{\rm L}=\Gamma_{\rm R}=\Gamma$), 
we perform a cycle operation where the levels of the dot 1 and 2 are modulated as 
\begin{align}
\varepsilon_1(t) &= \bar{E} + \varepsilon_{\rm rad} \cos(2\pi t/\tau_c), 
\\
\varepsilon_2(t) &= \bar{E} + \varepsilon_{\rm rad} \sin(2\pi t/\tau_c). 
\end{align}
For a slow modulation of this cycle, 
we calculate the pumped electron transfer $\langle \Delta q \rangle^{\rm ex}_{\tau_c}$ 
from the left reservoir to the double dot system by using the formula (\ref{average_ex}).

In Fig.~\ref{fig:Riwar}, we show the numerical results of $\langle \Delta q \rangle^{\rm ex}_{\tau_c}$ 
for various values of $U$ as a function of the cycle center $\bar{E}$. 
The vertical axis is normalized by $A \Gamma^{-2}$, 
where $A = \pi\varepsilon_{\rm rad}{}^2$ is the cycle area in the parameter space.
In Ref.~\onlinecite{RiwarSplettstoesser}, they showed 
that the positive resonant peaks appear around $\bar{E}=v$ 
and the negative resonant peaks appear around $\bar{E}=-v-U$ in our notation.
We can see that our numerical results are consistent with their results at this point.\cite{note:Riwar}

In Ref.~\onlinecite{RiwarSplettstoesser}, they also showed that 
the normalized electron transfer approaches $A$-independent value for small $A$.
We observe this behavior in the inset of Fig.~\ref{fig:Riwar}, 
where we plot the normalized electron transfer $\langle \Delta q \rangle^{\rm ex}_{\tau_c}/A \Gamma^{-2}$ 
at the negative peak for $\beta U=10$ (indicated by the arrow) as a function of $A$.

From these observations, we conclude that our scheme works well 
and provides the consistent results with those in Ref.~\onlinecite{RiwarSplettstoesser}.


\begin{thebibliography}{99}
\bibitem{note:Riwar}
There is also a discrepancy: 
Our result for $U=0$ is an odd function of $\bar{E}$ 
whereas their result in Fig.~3 of Ref.~\onlinecite{RiwarSplettstoesser} is not. 
From their discussion, the positive and negative peaks for $U=0$ should be located 
at symmetric positions from the origin of the horizontal axis.
We guess that the horizontal axis in their plot was accidentally shifted.
\end{thebibliography}
\end{document}